\documentclass[fontsize=11pt,a4paper,toc=flat]{arxiv-article}
\usepackage{amsmath}
\usepackage{hyperref}
\usepackage{cleveref}
\usepackage[most]{tcolorbox}
\usepackage{graphicx} 
\usepackage{amsmath,amsthm}

\title{Cylindrical Trap Dependence in the Unitary Fermi Gas}

\begin{document}
\author{%
	\begin{minipage}{.94\textwidth}
		\begin{center} \dosserif%
			{\small
           \textbf{Adèle~Le~Borgne}\textsuperscript{}
			}
		\end{center}
		\authorBlock{}{\dosserif{}%
			Albert Einstein Center for Fundamental Physics,\\
			Institute for Theoretical Physics, University of Bern,\\
			Sidlerstrasse 5, CH-3012 Bern, Switzerland}
	\end{minipage}
}
\date{}
\begin{titlepage}


\maketitle

	\thispagestyle{empty}

   \vfill

   \abstract{The unitary Fermi gas serves as a tunable realization of a strongly coupled CFT, making it a powerful system for probing universal quantum many-body phenomena. Precise measurement of its properties remains experimentally challenging: finite-temperature effects and spatial inhomogeneity introduced by external trapping potentials can significantly distort observables. Cylindrical trap geometries are commonly used in experiments. This note extends an existing theoretical framework to this geometry, deriving how the cylindrical confinement modifies the dynamic structure factor at zero temperature. These results provide a necessary correction for the interpretation of experimental spectra in trapped unitary gases.  }
	\vfill

\end{titlepage}
\tableofcontents

\section{Introduction}\label{Introduction}
Phase transitions are among the most fascinating phenomena in theoretical physics, 
marking the boundary between distinct states of matter. Of particular interest are 
continuous phase transitions, characterized by a diverging correlation length — 
fluctuations occur on all length scales simultaneously and the system loses any 
intrinsic scale. At such a critical point, the theory becomes scale invariant: 
physics looks the same regardless of the length scale at which it is probed.

Generally, scale invariance enhances to the full conformal symmetry~\cite{Delamotte_2016} 
— invariance under the full group of local angle-preserving transformations. 
This enlargement of the symmetry group places strong constraints 
on the structure of the theory, fixing the form of correlation functions and 
organizing the operator content into representations of the conformal algebra. The 
framework that systematically exploits these symmetries is Conformal Field Theory 
(CFT), which provides the natural language for describing the universal properties 
of critical systems~\cite{ginsparg1988appliedconformalfieldtheory}.

There are nonrelativistic counterparts of relativistic conformal field theories relevant to systems of interest in atomic and condensed matter physics. The nonrelativistic analog of the conformal algebra is the so-called Schrödinger algebra and many concepts can be transferred to nonrelativistic field theories. A particularly compelling realization of these ideas arises in the physics of the 
unitary Fermi gas — a two-component Fermi gas tuned to infinite scattering length 
via a Feshbach resonance~\cite{Giorgini_2008}. In this strongly coupled regime, the 
$s$-wave scattering length diverges, all microscopic length scales drop out, and the 
system sits at an interacting fixed point of the renormalization group. The unitary 
Fermi gas is therefore both scale and conformally invariant, making it a realization 
of a nonrelativistic CFT (NRCFT)~\cite{Nishida_2007}. It is experimentally 
accessible in ultracold atomic gases~\cite{Zwierlein_2004, Biss_2022, Weimer_2015, 
Mukherjee_2017, Patel_2020}, where the interaction strength can be tuned with 
remarkable precision via an external magnetic field.

Solving the unitary Fermi gas is a fundamentally non-perturbative problem. The 
absence of any small coupling parameter renders standard perturbative expansions 
unusable — the only intrinsic scale of the system is the density itself. Among 
the available analytical frameworks, Effective Field Theory (EFT) 
offers a systematic approach: by organizing the dynamics in terms of the symmetries of the 
problem — in particular general coordinate invariance and conformal invariance — 
and a low-energy momentum expansion, EFT provides a framework to describe universal properties of the system, such as the equation of state, while parameters like the Bertsch parameter $\xi$ must be determined from experiment or microscopic calculations and then used as inputs.
 At low energies, the relevant degrees of 
freedom are not the underlying fermions but rather the Goldstone bosons associated 
with the spontaneous breaking of the $U(1)$ particle-number symmetry in the 
superfluid phase — the phonon. The EFT is therefore a superfluid hydrodynamics 
built around this single gapless mode, whose leading-order action is fully fixed by 
symmetry and reproduces the results of Thomas-Fermi theory and superfluid 
hydrodynamics, while higher-order terms in the momentum expansion introduce low-energy coefficients (or Wilsonian coefficients) encoding the non-perturbative microscopic physics~\cite{Son_2006}.

A central quantity of interest is the phonon dispersion relation, which describes 
the acoustic excitation branch of the superfluid. In the absence of a confining 
potential, it takes the form~\cite{castin2025}
\begin{equation}
\label{eq:general-dispersion}
    q_0(q) = c_s q \left(1 + \frac{\gamma}{8c_s^2}q^2 
    + \mathcal{O}(q^4 \log q)\right),
\end{equation}
where $q_0$ is the energy, $q$ is the magnitude of the momentum transfer, $c_s$ 
is the speed of sound, and $\gamma$ is a parameter governing the deviation of the 
spectrum from linearity. Determining the sign of $\gamma$ is both an experimental 
and theoretical challenge of central importance: a convex branch ($\gamma > 0$) 
kinematically allows the three-phonon Beliaev-Landau process, whereas a concave 
branch ($\gamma < 0$) favors the four-phonon Landau-Khalatnikov process, with 
direct consequences for the relaxation mechanisms of the superfluid. The sign could be influenced by the trapping potential or the temperature \cite{castin2025,Castin_2024}.

In the presence of a trapping potential, the notion of a dispersion relation is not well defined: translation invariance is broken, and momentum is no longer a good quantum number. The appropriate object is instead the dynamic structure factor, and what one calls the dispersion relation should be understood as the curve along which it peaks — a point we return to in detail below.
Building on the methods developed in Ref.~\cite{beane2025trappingpotentialdependenceunitaryfermi}, this work extends their application to a trapping geometry that more closely reflects current experimental realizations of the unitary Fermi gas~\cite{Patel_2020}. We focus on the cylindrical geometry and provide a systematic treatment of the inhomogeneity induced by the confining potential.
As in the spherical case, the EFT exhibits a clear hierarchy of scales, and the corrections to the homogeneous result induced by the trapping potential organize themselves into three regimes with distinct power counting. This power counting combines the EFT derivative expansion with a Wentzel–Kramers–Brillouin (WKB) approximation, which amounts to an expansion in gradients of the trapping potential. In each regime, we obtain the energy spectrum by explicitly computing the phonon-field fluctuations, and extract the modifications to the dynamic structure factor induced by the corresponding density fluctuations.
The most interesting behavior occurs in the low-energy regime: the correction to the linear behavior depends on a combination of the longitudinal and transverse quantum numbers. For a transverse harmonic potential, the combination is
\begin{equation}
   q_0(\mathbf{q})=\sqrt{\frac{2\mu}{3}} \mathrm{q_\perp} \left(1+\left(\frac{n_z^2 \pi^2}{2\mathrm{~L}^2 }-\frac{5}{4 R^2_{cl,\rho}}\right)\frac{1}{q_\perp^2}\right),
\end{equation}
 with $R_{cl,\rho}$ the radial radius and $L/2$ the longitudinal radius. For the cloud dimensions realized in current experiments it is convex in general, but becomes concave in the special case of no longitudinal excitation $n_z=0$.

The remainder of the paper is organized as follows. In Section \ref{Sec:Experiments} and Section \ref{Sec:DSF}, we review the experimental trapping geometries and the Bragg spectroscopy techniques used to access the dynamic structure factor. Section \ref{Sec:Superfluid-EFT}, recalls the effective field theory of the unitary Fermi gas. In Section \ref{Sec:Cylinder geometry}, we derive the spectrum and dynamic structure factor in the cylindrical geometry and compare the result with experiment.

\section{Experimental setup and observables}\label{Sec:Experiments}
We compare our predictions primarily for the dispersion relation with the results of \cite{Biss_2022}. The system studied in that work was itself prepared using methods similar to those described in \cite{Mukherjee_2017} and \cite{Gaunt_2013}. We will briefly review here some trapping techniques.
Experiments on ultracold Fermi gases typically work with lithium-6 ($^6$Li) atoms, a fermionic isotope prepared in a balanced spin mixture of its two lowest hyperfine states, forming a two-component system in which s-wave interactions are permitted. 

The trapping technique depends on what one is interested in measuring. 
For momentum distribution, pair condensation, and the dispersion relation, it is more convenient to create a uniform trap that produces a spatially uniform density, in contrast to conventional harmonic traps that induce a spatially varying density that obscures critical phenomena. 
The radial confinement is achieved using blue-detuned light at 532 nm, which is repulsive for the atoms and therefore confines them in the interior of a hollow cylindrical beam. This beam is shaped by passing a collimated Gaussian beam through an axicon, a conical lens that converts it into a Bessel beam, which is subsequently focused through a microscope objective to produce a ring-shaped intensity pattern.
A circular opaque silver mask placed at the focal plane removes residual light from the interior of the ring and sharpens the potential wall.
The steepness of this radial wall is well described by a power-law potential $V(r) \propto r^m$ with $m = 16.2 \pm 1.6$, confirming that the confinement closely approximates a hard-wall box. 
Axial confinement is provided by two light sheets serving as end caps, generated from a second 532 nm beam detuned by 160 MHz to avoid interference, split into two elliptically shaped beams and projected onto the atoms through a rectangular mask.
Experiments aim for uniformity in all three directions: every region of the gas shares the same density, energy scale, and local chemical potential, realizing translational symmetry throughout the bulk. 
They claim that the distinction between a perfectly flat potential and $V(r) \propto  r^{16 \pm 2}$ to be irrelevant for most many-body studies.\\

Bragg spectroscopy \cite{Carcy_2019} is the most modern technique used to probe the dynamic structure factor
and excitation spectrum of ultracold quantum gases. It relies on the coherent
two-photon interaction between the atoms and two laser beams with slightly different
frequencies and wavevectors. When an atom absorbs a photon from one beam and is
stimulated to emit into the other, it receives a well-defined momentum kick
$\hbar \mathbf{q} = \hbar(\mathbf{k}_1 - \mathbf{k}_2)$ and a well-defined energy
transfer $\hbar\omega = \hbar(\omega_1 - \omega_2)$. By varying the frequency
difference $\omega_1 - \omega_2$ while keeping the momentum transfer fixed through the
angle between the beams, one can map out the excitation spectrum of the gas at a chosen
wavevector $\mathbf{q}$. The response of the system is captured by the dynamic structure factor
$S(\mathbf{q}, \omega)$, if one assumes isotropy then $S(\mathbf{q}, \omega)$=$S(q, \omega)$, which encodes how density fluctuations at wavevector $q$ evolve at
frequency $\omega$. In practice, the experiment measures the number of atoms transferred
to a different momentum state, or equivalently the energy deposited into the gas, as a
function of the frequency difference between the two Bragg beams. The resulting spectrum
reveals the position and width of collective modes such as the Bogoliubov phonon branch
in a Bose gas or the pair-breaking and phonon excitations in a strongly interacting
Fermi gas. In the linear response regime, where the Bragg pulse is weak enough not to
perturb the system significantly, the transferred energy is directly proportional to
$S(q,\omega)$, providing a clean spectroscopic window onto the many-body physics of
the gas. 
\section{Superfluid EFT to NLO}\label{Sec:Superfluid-EFT}

For completeness, we briefly review the EFT of Ref.~\cite{Son_2006}. Working at 
zero temperature and at unitarity, the dynamics of the Goldstone boson are governed 
by the Lagrangian
\begin{equation}\label{eq:eft-lag}
    \mathcal{L} = c_0 X^{5/2} + c_1 X^{-1/2}(\nabla X)^2 
    + c_2 X^{1/2} \left[(\Delta\theta)^2 
    - 3(\nabla\otimes\nabla\theta)^2\right] + \ldots,
\end{equation}
where the dots denote operators containing more derivatives, and
\begin{equation}\label{eq:X}
    X = \dot{\theta} - \frac{1}{2}(\nabla\theta)^2 - V.
\end{equation}
is a Galilean-invariant combination that serves as the fundamental building block 
of non-relativistic EFTs, with $V$ the external trapping potential and 
$\dot{\theta} \equiv \partial_t\theta$. The operators constituting the Lagrangian are the most general ones allowed at that order by the combined 
symmetries of the problem: general coordinate invariance, $U(1)$ symmetry, and 
conformal invariance. The low-energy constants $c_0,c_1,c_2$ encode the microscopic physics, they cannot be determined within the EFT; one needs lattice simulations or a large-N expansion to extract them.

\subsection*{Ground state and mean-field pressure}

The structure of the EFT is more transparent when one separates the classical 
ground state from the fluctuations. Decomposing the Goldstone field as
\begin{equation}\label{eq:decomp}
    \theta(t,\mathbf{r}) = \mu t + \pi(t,\mathbf{r}),
\end{equation}
where $\mu t$ is the time-dependent vacuum expectation value and 
$\pi(t,\mathbf{r})$ is the phonon fluctuation field, one finds that in the 
classical ground state the combination reduces to $X = \mu$. 
Evaluating the leading-order Lagrangian on this background directly yields the 
mean-field pressure,
\begin{equation}\label{eq:pressure-LO}
    P = c_0\,\mu^{5/2},
\end{equation}
whose form $P \propto \mu^{d/2+1}$ is entirely fixed by conformal invariance. The 
coefficient $c_0$ is related to the Bertsch parameter $\xi$ by
\begin{equation}\label{eq:c0}
    c_0 = \frac{2^{5/2}}{15\pi^2\,\xi^{3/2}},
\end{equation}
encoding all the non-perturbative physics of the strongly coupled ground state 
into a single universal constant that must be determined experimentally or by 
quantum Monte Carlo simulations. The Bertsch parameter can be understood as the ratio of ground-state energy of a strongly interacting Fermi gas to the energy of a completely non-interacting gas.
 "Universal" means it does not depend on the type of fermion or the specific details of how they interact, as long as they are at the unitary limit of maximum interaction strength.

\subsection*{Phonon fluctuations and dispersion relation}

Expanding around the ground state~\eqref{eq:decomp} and retaining terms up to quadratic order
in $\pi(t,\mathbf{r})$, one finds, at leading order in the EFT, that the phonon propagates with a linear dispersion relation,
\begin{equation}\label{eq:dispersion-LO}
    q_0 (\mathbf{q})= c_s q, \qquad c_s = \sqrt{\frac{\xi}{3}}\,v_F,
\end{equation}
where $\,|\mathbf{q}|=q$ and $v_F=\hbar k_F/m$ is the Fermi velocity. This linearity is a direct consequence of the nature of the Goldstone mode. The NLO operators proportional to $c_1$ and $c_2$, 
evaluated as tree-level insertions on the classical background for a homogeneous system ($V = 0$), correct the 
dispersion relation as~\cite{Son_2006}
\begin{equation}\label{eq:dispersion-NLO}
    q_0(\mathbf{q}) = c_s\,q\left[1 - \pi^2\sqrt{2\xi}
    \left(c_1 + \frac{3}{2}c_2\right)\frac{q^2}{k_F^2}\right] 
    + \mathcal{O}(q^5\ln q).
\end{equation}
Since the number of independent physical observables depending on $c_1$ and $c_2$ 
— including the dispersion relation, the density response function, and the 
ground-state energy in a trap — exceeds two, the EFT remains predictive at NLO despite the non-perturbative nature of the problem.

\subsection*{NLO terms versus loops}

A natural question is whether quantum loop corrections must be included alongside the NLO tree-level terms. Son and Wingate 
demonstrate that this is not the case~\cite{Son_2006}: loop contributions are 
parametrically suppressed and enter only at next-to-next-to-leading order (NNLO). 
This follows from a simple power counting argument. Working in units where $c_s=1$, 
the Lagrangian takes the schematic form
\begin{equation}\label{eq:power-counting}
    \mathcal{L} \sim (\partial\theta)^2 
    + \frac{\mathcal{O}(1)}{\mu^2}(\partial_0\theta)(\partial_i\theta)^2 
    + \frac{\mathcal{O}(1)}{\mu^4}(\partial\theta)^4 + \ldots,
\end{equation}
where the NLO tree-level terms are suppressed by $\mu^{-2}$ relative to the 
leading-order kinetic term, i.e., they are of order $\mathcal{O}(p^2)$. A 
one-loop correction to the phonon self-energy, by contrast, involves two 
three-phonon vertices each carrying a factor of $\mu^{-2}$, making the loop 
contribution proportional to $\mu^{-4}$ and therefore of order 
$\mathcal{O}(p^4)$. This loop diagram is furthermore UV divergent and requires 
regularization, after which it produces a finite contribution with a possible 
logarithmic singularity in the external momentum; the UV divergence is absorbed 
into a renormalization of NNLO Wilson coefficients, consistently with the power 
counting. The conclusion is therefore that, at leading $\mathcal{O}(1)$ and next-to-leading $\mathcal{O}(p^2)$ order, the EFT is entirely classical and loop corrections can be consistently neglected.

In this work, we generalize the NLO dispersion 
relation~\eqref{eq:dispersion-NLO} to the case of a non-homogeneous system in the 
presence of a trapping potential. The spherically symmetric case has been treated 
in~\cite{beane2025trappingpotentialdependenceunitaryfermi}; here we focus on 
the cylindrical geometry, which more closely reflects current experimental 
realizations of the unitary Fermi gas~\cite{Patel_2020}.
\section{Cylinder geometry}\label{Sec:Cylinder geometry}
We consider the potential
\begin{equation}
    V(\textbf{r})=V^{\parallel}(z)+V^{\perp} (r), \quad V^{\parallel}(z)= \left\{
    \begin{array}{ll}
        0& \mbox{if } \mid z\mid < L/2 \\
        \infty & \mbox{if } \mid z\mid > L/2 
    \end{array}
\right. \quad  \text{and } V^{\perp} (r)=\mu \left(\frac{r}{R_{cl,\rho}}\right)^{2k}.
\end{equation}
The original cylindrical coordinates are $(r, \vartheta, z)$, where $r$ is the transverse 
radial coordinate, $\vartheta$ the azimuthal angle, and $z$ the longitudinal coordinate.
\begin{figure}[h!]
    \centering
    \includegraphics[width=1\linewidth]{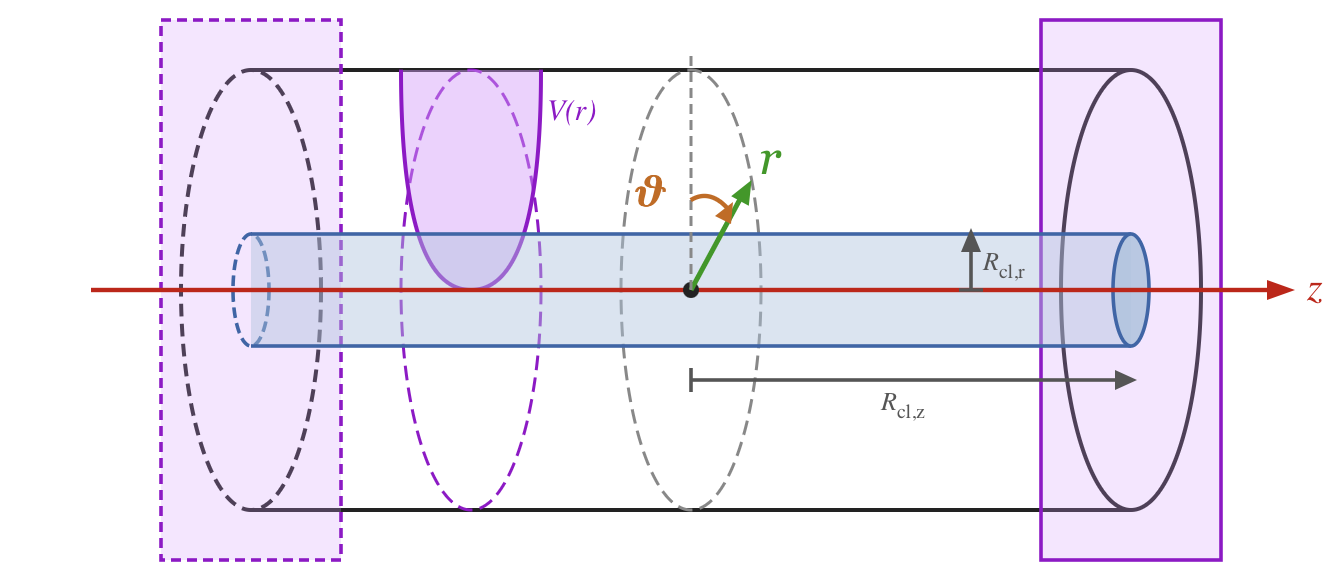}
    \caption{Condensate confined in cylinder-shape geometry, via a power law potential in the transverse direction and by end caps in the longitudinal direction. There is no potential along $z$; the solutions are plane waves with frequency depending on the boundary conditions.}
\end{figure}

\subsection{Scales}
We perform the same scale analysis as in \cite{beane2025trappingpotentialdependenceunitaryfermi}.
\paragraph{The small parameter \(\iota\).}
The boundary of the cloud is defined as the distance from the origin at which 
the density $n(\mathbf{r})=\frac{\partial \mathcal{L}}{\partial \mu}$, vanishes in the LO EFT,
\begin{equation}
    \mu-V(\textbf{r})=0.
\end{equation}
In the cylinder geometry there are two distinct cloud scales: the transverse cloud 
radius $R_{cl,\rho}$, defined by $\mu - V^\perp(R_{cl,\rho}) = 0$, and the longitudinal 
half-length $R_{cl,z} = L/2$. We introduce the 
corresponding dimensionless rescaled coordinates
\begin{equation}
    \rho=\frac{r}{R_{cl,\rho}}, \quad u_z=\frac{z}{R_{cl,z}},
\end{equation}
where $\theta$ is shared between the original and rescaled systems since it is already 
dimensionless. We also rescale the potential as
\begin{equation}
    V=\mu \bar V.
\end{equation}
The action at the saddle is
\begin{equation}
\begin{aligned}
     \mathcal{L}
     &= \mu^{5/2}\left( c_0(1  - \bar{V}  ) ^{5/2}+\frac{c_1}{R^2_{cl,\rho}\mu}(1 -\bar{V})^{-1/2}(\frac{d}{d \rho} \bar{V})^2\right).\\
\end{aligned}
\end{equation}
As in the spherically symmetric case, 
the subleading term is controlled by a small parameter, here denoted by $\iota$,
\begin{equation}
    \iota^2=\frac{2}{R^2_{cl,\rho}\mu}.
\end{equation}
It is convenient to rewrite $\iota$ as the ratio of a low-energy and a high-energy scale, 
introducing $\varpi$ such that
\begin{align}
  \varpi &= \frac{\sqrt{2\mu}}{R_{cl,\rho}}  &\text{and} &&  \iota &= \frac{\varpi}{\mu } \, .
\end{align}
\paragraph{The small parameter \(\eta\).}

Expanding the action up to second order in the fluctuation field $\pi(t,\mathbf{r})$ we obtain the 
quadratic action encoding the small fluctuations:
\begin{multline}
\label{eq:quadratic-action}
\mathcal{L}^{(2)}[\pi] = -\frac{5}{8}c_0 \sqrt{\mu-V} \pqty*{2(\mu-V)(\nabla \pi)^2-3\dot\pi^2} \\
+ c_1 \bqty*{\frac{(\nabla \dot \pi)^2 }{\sqrt{\mu-V}}  + \frac{\nabla V \cdot \nabla \dot \pi^2}{2(\mu-V)^{3/2}} + \frac{(\nabla V)^2\pqty*{2(\mu-V)(\nabla \pi)^2+3\dot\pi^2}}{8(\mu-V)^{5/2}}} \\
+  c_2\sqrt{\mu-V} \pqty*{(\Delta \pi)^2 -3 (\nabla \otimes \nabla \pi)^2}.
\end{multline}
We introduce the typical energy of the fluctuations $q_0$, which we use to rescale the time variable:
\begin{equation}
	\bar t = q_0 t.
\end{equation}
Expressed in terms of the dimensionless coordinates $(\rho, \vartheta, u_z)$, where 
$\perp$ (resp.\ $\parallel$) denotes derivatives with respect to the transverse 
coordinates $(\rho, \vartheta)$ (resp.\ $u_z$), the action takes the form:
\begin{multline}
\mathcal{L}^{(2)}[\pi] = -\frac{5}{8}c_0 \frac{\mu^{3/2} }{R^2_{cl,\rho}}\sqrt{1 - \bar V} \pqty*{2  (1 - \bar V)(\nabla^{\perp}\pi+\frac{R_{cl,\rho}}{R_{cl,z}}\nabla^{\parallel} \pi)^2 - 3 \frac{q_{0}^2 R_{cl,\rho}^2}{\mu} \pqty*{\pdv{\pi}{\bar t}}^2} \\
+  c_2 \frac{\mu^{1/2}}{R_{cl,\rho}^4} \sqrt{1 - \bar V} \bqty*{(\Delta^{\perp} \pi+\frac{R^2_{cl,\rho}}{R^2_{cl,z}}\Delta^{\parallel} \pi)^2 -3 \pqty*{(\nabla^{\perp}+\frac{R_{cl,\rho}}{R_{cl,z}}\nabla^{\parallel} ) \otimes (\nabla^{\perp}+\frac{R_{cl,\rho}}{R_{cl,z}}\nabla^{\parallel} )} \pi}^2 \\
+ c_1 \frac{\mu^{1/2}}{R_{cl,\rho}^4} \Bigg[\frac{(\nabla^{\perp} \bar V )^2}{4(1 - \bar V)^{3/2}} (\nabla^{\perp}\pi+\frac{R_{cl,\rho}}{R_{cl,z}}\nabla^{\parallel}  \pi)^2  +  \frac{q_{0}^2 R_{cl,\rho}^2}{\mu} \Bigg(\frac{((\nabla^{\perp}+\frac{R_{cl,\rho}}{R_{cl,z}}\nabla^{\parallel} )\pdv{\pi}/{\bar t})^2 }{\sqrt{1 - \bar V}}\\
+ \frac{\nabla^{\perp} \bar V \cdot (\nabla^{\perp}+\frac{R_{cl,\rho}}{R_{cl,z}}\nabla^{\parallel} )(\pdv{\pi}/{\bar t})^2}{2( 1 - \bar V)^{3/2}} +  \frac{3 \pqty*{\pdv{\pi}/{\bar t} \nabla^{\perp} \bar V }^2 }{8(1 - \bar V)^{5/2}}) \Bigg].
\end{multline}
The longitudinal derivatives come with the ratio $\epsilon = R_{cl,\rho}/R_{cl,z}$, which reflects the geometry of the cylinder, we will not assume that this ratio is small nor large. The expansion is consistent if at leading order $q_0 \sim \sqrt{\mu}/R_{cl,\rho}$, and the controlling parameter is
\begin{equation}
	\eta = \frac{q_0}{\mu}.
\end{equation}
Then the leading term is controlled by the low-energy constant $c_0$ and the NLO 
correction is proportional to $c_1$ and $c_2$, which contribute at the same order.

\paragraph{The small parameter \(\delta\).}

The equation of motions of the LO quadratic action are
\begin{equation}
   \mathcal{L}^{(2)}_{c_0}= -\frac{5}{8}c_0 \frac{\mu^{3/2} }{R^2_{cl,\rho}}\sqrt{1 - \bar V} \pqty*{2  (1 - \bar V)\left((\nabla^{\perp}\pi)^2+\epsilon^2(\nabla^{\parallel} \pi)^2\right) - 3 \frac{q_{0}^2 R_{cl,\rho}^2}{\mu} \pqty*{\pdv{\pi}{\bar t}}^2}.
\end{equation}
Since the system is time-translation invariant, we can write the solution as
\begin{equation}
	\pi(t, \boldsymbol{\rho}) = e^{iq_0t} \pi(\boldsymbol{\rho}).
\end{equation}
This reduces the equation of motion to
\begin{align}
 \frac{2 \mu}{q_0^2 R^2_{cl,\rho}} \left( \nabla^{\perp}((1 - \bar V)^{3/2}\nabla^{\perp} \pi(\boldsymbol{\rho}))+\epsilon^2\nabla^{\parallel}( (1 - \bar V)^{3/2}\nabla^{\parallel} \pi(\boldsymbol{\rho}))\right)=   -3 \sqrt{1 - \bar V}  \pi(\boldsymbol{\rho}).
\label{linear-eom}
\end{align}
The leading derivative in this equation is multiplied by the coefficient $\delta^2$, where
\begin{equation}
    \delta^2= \frac{2\mu}{q_0^2 R^2_{cl,\rho}} .
\end{equation}
In the limit $\delta \to 0$, the equation of motion lends itself to a WKB expansion. 
The three scales $\varpi$, $q_0$, and $\mu$ are the same as in the spherically symmetric 
case, with the replacement 
$R_{cl} \to R_{cl,\rho}$, and they define the same two expansion parameters: $\eta = q_0/\mu$ 
controlling the EFT expansion, and $\delta = \varpi/q_0$ controlling the WKB expansion. 
Assuming scale separation $\varpi \ll q_0 \ll \mu$, we can distinguish three regimes 
depending on the relative size of $\eta$ and $\delta$:
\begin{itemize}
  \item \textbf{Low-energy regime} ($\varpi \ll q_0 \ll \sqrt{\varpi\mu}$, i.e.\ 
  $\eta \ll \delta \ll 1$): $\delta$ is not small enough for the physical optics 
  approximation to suffice, so higher-order WKB corrections are needed; however 
  $\eta$ is negligible and the LO EFT is sufficient.
   \item \textbf{Linear regime} ($q_0 \approx \sqrt{\mu \varpi}$): both $\delta$ and $\eta$ are 
  small, the dynamics is well described by the LO EFT and the LO WKB (physical optics) 
  approximation.
   \item \textbf{High-energy regime} ($\sqrt{\varpi\mu} \ll q_0 \ll \mu$, i.e.\ 
  $\delta \ll \eta \ll 1$): $\delta$ is small so the physical optics approximation 
  is valid, but $\eta$ is no longer negligible and NLO EFT corrections must be included.
\end{itemize}
We now determine the energy spectrum and the phonon field in each of the three regimes.
\subsection{Linear regime (LO EFT -- LO WKB)}
We can solve (\ref{linear-eom}) with the ansatz
\begin{align}
	\pi(\boldsymbol{\rho}) = 
\sin(k_z (z +\frac{L}{2})) e^{in_\vartheta\vartheta}\pi(\rho), \quad k_z=\frac{ \pi n_z }{L},
\end{align}
with $n_z$ the longitudinal quantum number, and $n_\theta$ the azimuthal quantum number.
The radial part is
\begin{align}
 \delta^2 \left( \nabla^{\perp}((1 - \bar V)^{3/2}\nabla^\perp \pi(\boldsymbol{\rho}))+\epsilon^2\nabla^\parallel( (1 - \bar V)^{3/2}\nabla^{\parallel} \pi(\boldsymbol{\rho}))\right)&=   -3 \sqrt{1 - \bar V}  \pi(\boldsymbol{\rho}),\\
 \delta^2 \left( \frac{\partial}{\partial \rho}\left(\rho (1 - \bar V)^{3/2}\frac{\partial}{\partial \rho}\pi(\rho)\right)-\left(\frac{n_\theta^2}{\rho} +\frac{(\pi n_z\epsilon)^2}{4}\rho\right)(1 - \bar V)^{3/2} \pi(\rho)\right)&=-3\rho \sqrt{1 - \bar V}  \pi(\rho). \label{equ:LOEFT-eom}
\end{align}
The angular part is now $-\left(\frac{n_\theta^2}{\rho^2} +\frac{(\pi n_z\epsilon)^2}{4}\right)$ instead of $\frac{-\ell(\ell+1)}{u^2}$ as in the spherical case. We can solve the radial equation of motion using the WKB ansatz,

\begin{equation}
    \pi(\rho)=e^{\frac{i}{\delta}S_0(\rho)+S_1(\rho)+i \delta S_2(\rho)+ \delta^2 S_3(\rho)+O(\delta^4)}.
\end{equation} 
The even subscripts capture the change in phase and the odd ones capture the change in the amplitude. The \textit{eikonal} ($\sim \frac{1}{\delta}$) and \textit{transport} ($\sim \delta^0$) equations are
\begin{align}
   &(1-\bar V(\rho))S_0'(\rho)^2 =3, \\
    &S_0'(\rho)\left[3\rho\bar V'(\rho)-2(1-\bar V(\rho))\right]-2\rho(1-\bar V(\rho))\left[S_0''(\rho) +2 S_0'(\rho) S_1'(\rho)\right]=0.
\end{align}
This leads to,
\begin{align}
  S_0(\rho) &= \pm \sqrt{3}  \int_{\rho_0}^\rho \frac{\dd{w}}{\sqrt{ 1 - \bar V(w)}}\,,\label{equ:linear-sol} \\
  S_1(\rho) &= - \log( \sqrt{\rho( 1 - \bar V(\rho))}) + k_1 .
\end{align}
Requiring regularity at the center of the cylinder $\rho=0$ and at the edge of the cloud $ \rho=1$, we obtain in the case of a power law potential $\bar V(\rho)=\rho^{2k}$ the normalized field
\begin{equation}\label{eq:physical-optics-superharmonic-oscillator}
\pi(\rho) = \sqrt{2k \binom{\sfrac{1}{2k} - \frac{1}{2}}{\frac{1}{2}}} \frac{ \sin\left( \frac{2 n}{\binom{1/(2k)}{1/2}} \rho \  _2F_1(\frac{1}{2},\frac{1}{2k},1+\frac{1}{2k};\rho^{2k} )\right)}{ \sqrt{\rho(1 - \rho^{2k})}} ,
\end{equation}
with spectrum, 
\begin{equation}
\mathrm{q}_0(n)=\frac{2 n}{\sqrt{3}\binom{1 /(2 \mathrm{k})}{1 / 2}} \varpi ,
\end{equation}
where $n$ is the radial quantum number $n=0,1,2,...$.
The solution in the linear regime is almost identical to the one in the spherical case.
\subsection{Low-energy regime (LO EFT -- NLO WKB)}
To go beyond the linear approximation in the low-energy regime, we need higher-order 
corrections to the WKB expansion. 

As in the spherical case, the centrifugal term $\frac{1}{\rho^2}$ introduces a singular point at the origin.
Naive WKB, which is constructed to handle simple turning points, fails to reproduce the correct  $\rho^{|n_\theta|}$  behavior of the wavefunction near $\rho = 0$.
Following the Langer modification \cite{Kramer_2010,Grossmann_2011}, the net effect is implemented directly at the level of the WKB integrand, by the replacement $n_\theta^2  \rightarrow n_\theta^{2'}= n_\theta^2 + 1/4$ in the effective centrifugal coefficient.
After this modification, we restrict to the physical rotationally-invariant modes $n_\theta = 0$ which corresponds to the new minimal value $n_\theta' =  1/2$. In the following, the Langer shift is implicit and we will refer to $n_\theta'=1/2$ as the s-wave, the lowest rotational mode.
%
%
%
%
The equations to solve are
\begin{align}
&S_1'(\rho)\left(3\rho\bar V'(\rho)-2(1-\bar V(\rho))\right)-2\rho(1-\bar V(\rho))(S_1'(\rho)^2-2S_0'(\rho)S_2'(\rho)+S_1''(\rho))\\
&\hspace{7cm}+(\frac{1+4 n_\theta^2}{2\rho}+\frac{(\pi n_z \epsilon)^2}{2} \rho)(1-\bar V(\rho))=0,\\
&S_2'(\rho)\left(3\rho \bar V'(\rho)-2(1-\bar V(\rho))\right)-2\rho(1-\bar V(\rho))\left(2S_1'(\rho)S_2'(\rho)+2S_0'(\rho)S_3'(\rho)+S_2''(\rho)\right)=0.
\end{align}
this leads to
\begin{align}
   S_2(\rho)&= \int_{\rho_0}^\rho\frac{1}{4\sqrt{3-3\bar V(w)}} \underbrace{\left(\bar V''(w)+\frac{3\bar V'(w)}{2w }-(2\frac{ n_\theta^2}{w^2}+\frac{(\pi n_z\epsilon)^2}{2})(1-\bar V(w))\right)}_{F_{n_\theta,n_z}(w)} dw,\\
    S_3(\rho)&=\frac{-1}{24}\left(\bar V''(\rho)+\frac{3\bar V'(\rho)}{2\rho}-\frac{2 n_\theta^2}{\rho^2}(1-\bar V(\rho))+\frac{(\pi n_z \epsilon)^2}{2}\bar V(\rho)\right).
\end{align}
We expect that upon summation of the full WKB expansion the singularity 
at the origin will cancel, as we will see for the harmonic potential an exact 
solution exists and does not contain singularities. In the low energy regime, we will therefore restrict the analysis to the s-wave. At this order, we find
\begin{equation}
  S_1(\rho) + \frac{\varpi^2}{q_0^2} S_3(\rho) = - \frac{1}{2}\log(\rho) - \frac{1}{2} \log*( 1- \bar V(\rho)) - \frac{\varpi^2}{24 q_0^2} F_{0,n_z}(\rho) + \order*{\frac{\varpi^4}{q_0^4}} \, ,
\end{equation}
that we can rewrite as
\begin{equation}
  S_1(\rho) + \frac{\varpi^2}{q_0^2} S_3(\rho) = - \frac{1}{2} \log*(\rho + \frac{\rho \varpi^2}{12 q_0^2} F_{0,n_z}(\rho)) - \frac{1}{2} \log( 1- \bar V(\rho)) + \order*{\frac{\varpi^4}{q_0^4}} \, .
\end{equation}
Imposing regularity at $\rho = 0$, we find that the solution must take the form
\begin{equation}
  \pi(\rho)   = D \frac{\sin*( \frac{q_0}{\varpi} \sqrt{3} \int_0^\rho \frac{\dd{w}}{\sqrt{1 - \bar V(w)}} + \frac{\varpi}{q_0} \frac{1}{4\sqrt{3}} \int_0^\rho \frac{F_{0,n_z}(w)}{\sqrt{1 - \bar V(w)} } \dd{w}  )}{\sqrt{\rho\pqty*{1 + \frac{\varpi^2}{12 q_0^2}F_{0,n_z}(\rho) } \pqty*{1 - \bar V(\rho)}}}  \,.
\end{equation}
Imposing regularity at $\rho = 1$ yields the quantization condition with,
\begin{align}
	\frac{q_0}{\varpi} \sqrt{3} \int_0^1 \frac{\dd{w}}{\sqrt{1 - \bar V(w)}} + \frac{\varpi}{q_0} \frac{1}{4 \sqrt{3}} \int_0^1 \frac{F_{0,n_z}( w)}{\sqrt{1 - \bar V(w)} } \dd{w} &= n \pi\, , & n&=0,1,2,3,\dots
\end{align}

For a power-law potential \(\bar V(\rho)  = \rho^{2k}\) it is possible to solve the integrals analytically to find
\begin{equation}
 \frac{q_0(n)}{\varpi} = \sqrt{\frac{\pi}{3}}\,\frac{\Gamma\!\left(\frac{1+k}{2k}\right)}{\Gamma\!\left(1 + \frac{1}{2k}\right)}\, n 
- \frac{1}{16\sqrt{3\pi}\,n}\left[\frac{2(1+4k)\,\Gamma\!\left(1 - \frac{1}{2k}\right)}{\Gamma\!\left(\frac{3}{2} - \frac{1}{2k}\right)} 
- \frac{ n_z^2\,\pi^2 \epsilon^2\,\Gamma\!\left(1 + \frac{1}{2k}\right)}{\Gamma\!\left(\frac{1}{2}\!\left(3 + \frac{1}{k}\right)\right)}\right] + \cdots
\end{equation}
In the case of a transverse harmonic potential \(k = 1\), this reduces to
\begin{equation}\label{equ:low-energy-spectrum}
  \frac{q_0 (n)}{\omega} = \frac{2}{\sqrt{3}} n +
  \frac{-20+(\pi n_z \epsilon)^2}{32 \sqrt{3}} \frac{1}{n} +\dots
\end{equation}
\subsection{Transverse harmonic potential}
We found in the spherical case that the harmonic potential is exactly solvable. Here it is the case for the $s$-wave with some subtleties.
We can solve (\ref{equ:LOEFT-eom}) exactly for the harmonic potential $\bar V(\rho)=\rho^2$,
\begin{equation}\label{equ:harmonic-eom}
     \delta^2 \left( \frac{\partial}{\partial \rho}\left(\rho (1 - \rho^2)^{3/2}\frac{\partial}{\partial \rho}\pi\right)-\left(\frac{n_\theta^{2'}}{\rho} +\rho\frac{(\pi n_z\epsilon)^2}{4}\right)(1 - \rho^2)^{3/2} \pi\right)=-3\rho \sqrt{1 - \rho^2}  \pi(\rho).
\end{equation}
After the change of variable, $\pi(\rho)=\rho^{n_\theta '}\phi(\rho)$ and $\rho^2=x$, we have the following radial equation of motion:
\begin{align}
    \phi''(x)x(x-1)+\phi'(x)\left(-1-n_\theta'+(n_\theta' +\frac{5}{2})x\right)+\phi(x)\left(\frac{3}{4}n_\theta' -\frac{3}{4\delta^2}-\frac{(n_z\pi\epsilon)^2 }{16}x +\frac{(n_z\pi\epsilon)^2 }{16}\right)=0.
\end{align}
\begin{itemize}
    \item For the special case of $n_z=0$, the equation reduces to,
    \begin{align}
    \phi''(x)x(x-1)+\phi'(x)\left(-1-n_\theta'+(n_\theta' +\frac{5}{2})x\right)+\phi(x)\left(\frac{3}{4}n_\theta' -\frac{3}{4\delta^2}\right)=0,
    \end{align}
    which is exactly solvable in terms of hypergeometric function;
    \begin{equation}
        \phi(\rho)=\  _2F_1\left(\frac{3}{4}+\frac{n_\theta'}{2}-\frac{\sqrt{12 \delta^2+9 \delta^4+4 n_\theta^{2'}\delta^4}}{4\delta^2},\frac{3}{4}+\frac{n_\theta'}{2}+\frac{\sqrt{12 \delta^2+9 \delta^4+4n_\theta^{2'}\delta^4}}{4\delta^2},1+n_\theta'; \rho^2\right).
    \end{equation}
    The hypergeometric function is defined for $\abs{x} < 1$ by the power series around $z=0$, asking for the series to terminate at a degree $n$ insure regularity at $z=1$. This provides conditions on the coefficients of the series, and in particular a quantization condition on $\frac{1}{\delta}$,
    \begin{align}
        \delta^2&=\frac{3}{4n^2+4n  n_\theta'+3n_\theta'+6n}.
    \end{align}
    At large-n evaluating to the lowest $n_\theta'$ mode, 
    \begin{equation}
         \frac{q_0}{\omega}
        \approx \frac{2(n+1)}{\sqrt{3}}-\frac{5}{8\sqrt{3}n}+...
    \end{equation}
    This matches our previous result (\ref{equ:low-energy-spectrum}), with the shift $n\rightarrow n+1$ in WKB, as the matching is done at large-n, the identification $n\sim n+1$ is justified.
    \item For $n_z\ne 0$, there is an additional term $\frac{-(n_z\pi\epsilon)^2 }{16}x$ that prevents the equation from being a hypergeometric differential equation. However, we can rewrite it as
    \begin{align}
    \phi''(x)+\phi'(x)\left(\frac{3/2}{(x-1)}+\frac{1+n_\theta'}{x} \right)+\phi(x)\left(\frac{-\frac{(n_z\pi\epsilon)^2}{16} x+\frac{3}{4}n_\theta' -\frac{3}{4\delta^2} +\frac{(n_z\pi\epsilon)^2}{16}}{x(x-1)}\right)&=0,\\
     \phi''(x)+\phi'(x)\left(\frac{3/2}{(x-1)}+\frac{3/2}{x} \right)+\phi(x)\left(\frac{-\frac{(n_z\pi\epsilon)^2}{16} x+\frac{3}{8} -\frac{3}{4\delta^2} +\frac{(n_z\pi\epsilon)^2}{16}}{x(x-1)}\right)&=0 \label{equ:Heun-lowest-mode}.
    \end{align}
In (\ref{equ:Heun-lowest-mode}), we have set $n_\theta'$ to its minimal value in order to compare with the WKB approach. This is the confluent Heun's differential equation.
\begin{equation}
    \phi''(x)+(\alpha+\frac{\beta+1}{x}+\frac{\gamma+1}{x-1})\phi'(x)+(\frac{\mu}{x}+\frac{\nu}{x-1})\phi(x)=0
\end{equation}
There are three singular points, at $x=0,x=1$ and $x=\infty$. The standard confluent Heun's function $HeunC(\alpha, \beta,\gamma,\delta_H,\eta,x)$ is regular around $x=0$ and is defined on the unit disk $\abs{x}<1$ by
\begin{equation}\label{equ:Heun-series}
    HeunC(\alpha, \beta,\gamma,\delta_H,\eta,x)=\sum_{k=0}^\infty c_k x^k,
\end{equation}
with the relations $\mu=\frac{1}{2}(\alpha-\beta-\gamma+\alpha \beta -\beta \gamma)-\eta$ and $\nu=\frac{1}{2}(\alpha+\beta+\gamma+\alpha \gamma+\beta \gamma)+\delta_H+\eta$.
In this case, requiring the Heun series to terminate at finite order forces $n_z = 0$, so for $n_z \ne 0$, one must instead use the continued-fraction approach.
\begin{align}
&\phi''(x)x(x-1)+\phi'(x)\left(-3/2+3x\right)+\phi(x)\left(\frac{3}{8}-\frac{3}{4\delta^2}-\frac{(n_z\pi\epsilon)^2 }{16}x +\frac{(n_z\pi\epsilon)^2}{16}\right)=0,\\
&\left[(k+1)(k+\frac{3}{2}) \right]c_{k+1}=\left[k^2+2k +\frac{3}{8}-\frac{3}{4\delta^2} +\frac{(n_z\pi\epsilon)^2}{16} \right] c_k  -\frac{(n_z\pi\epsilon)^2 }{16}c_{k-1}. \label{equ:c_k1}
\end{align}
For the series (\ref{equ:Heun-series}) to converge at $x = 1$, the recurrence relation must admit a minimal solution. Equivalently, by Pincherle’s theorem \cite{AHLBRANDT1996188}, this condition is satisfied if and only if the associated continued fraction converges. This type of problem has been studied in the context of black hole physics, particularly in the computation of quasinormal modes and frequencies of charged, static black holes \cite{Leaver:1990zz,Onozawa_1996}.
We follow a similar numerical approach to determine the quantization condition of $\frac{1}{\delta}$. To this end, we first rewrite the recurrence relation in the form of a tridiagonal matrix,
\begin{equation}
\mathcal{M}\cdot\mathbf{c}=\begin{pmatrix}
b_0 & a_0 & 0   & 0   & \cdots \\
d_1 & b_1 & a_1 & 0   & \cdots \\
0   & d_2 & b_2 & a_2 & \cdots \\
0   & 0   & d_3 & b_3 & \cdots \\
\vdots & \vdots & \vdots & \vdots & \ddots
\end{pmatrix}
\begin{pmatrix}
c_0 \\ c_1 \\ c_2 \\ c_3 \\ \vdots
\end{pmatrix}=\frac{3}{4\delta^2}\mathbb{I} \cdot \mathbf{c},
\end{equation}
with $a_k= -(k+1)(k+\frac{3}{2}), b_k=k^2+2k +\frac{3}{8} +\frac{(n_z\pi\epsilon)^2}{16}, d_k= -\frac{(n_z\pi\epsilon)^2 }{16}$.
It reduces to a matrix eigenvalue problem, where the eigenvalues of $\mathcal{M}$ correspond to the discretized values of \( \frac{3}{4\delta^2} \). This problem is solved numerically by diagonalizing a matrix whose dimension is chosen to be much larger than the number of desired eigenvalues, ensuring that the states are not affected by finite-size truncation.

However, the lowest-lying states are not expected to match the WKB spectrum perfectly, since the latter is derived in the large-\( n \) limit. Ordering the eigenvalues from smallest to largest, we extract the first 200 levels of the spectrum. The results are shown in Fig.~\ref{fig:spectrum-verification}, where we fit the numerical eigenvalues to the following ansatz: $\frac{1}{\delta}=\frac{q_0}{\omega}=\frac{2}{\sqrt{3}}n+\frac{\alpha}{n}+\frac{\beta}{n^2}+\frac{\gamma}{n^3}+O(\frac{1}{n^4})$.
We find excellent agreement between the WKB approximation and the diagonalization of the tridiagonal matrix, with a difference at less than $10^{-4}\%$ in the $\alpha$ coefficient. 
 The larger the value of \( n_z \), the later this agreement sets in, improvement of the result can be achieved by adding more eigenvalues. 
\begin{figure}[ht]
  \centering
  \begin{minipage}[c]{0.7\textwidth}
    \centering
    \includegraphics[width=\linewidth]{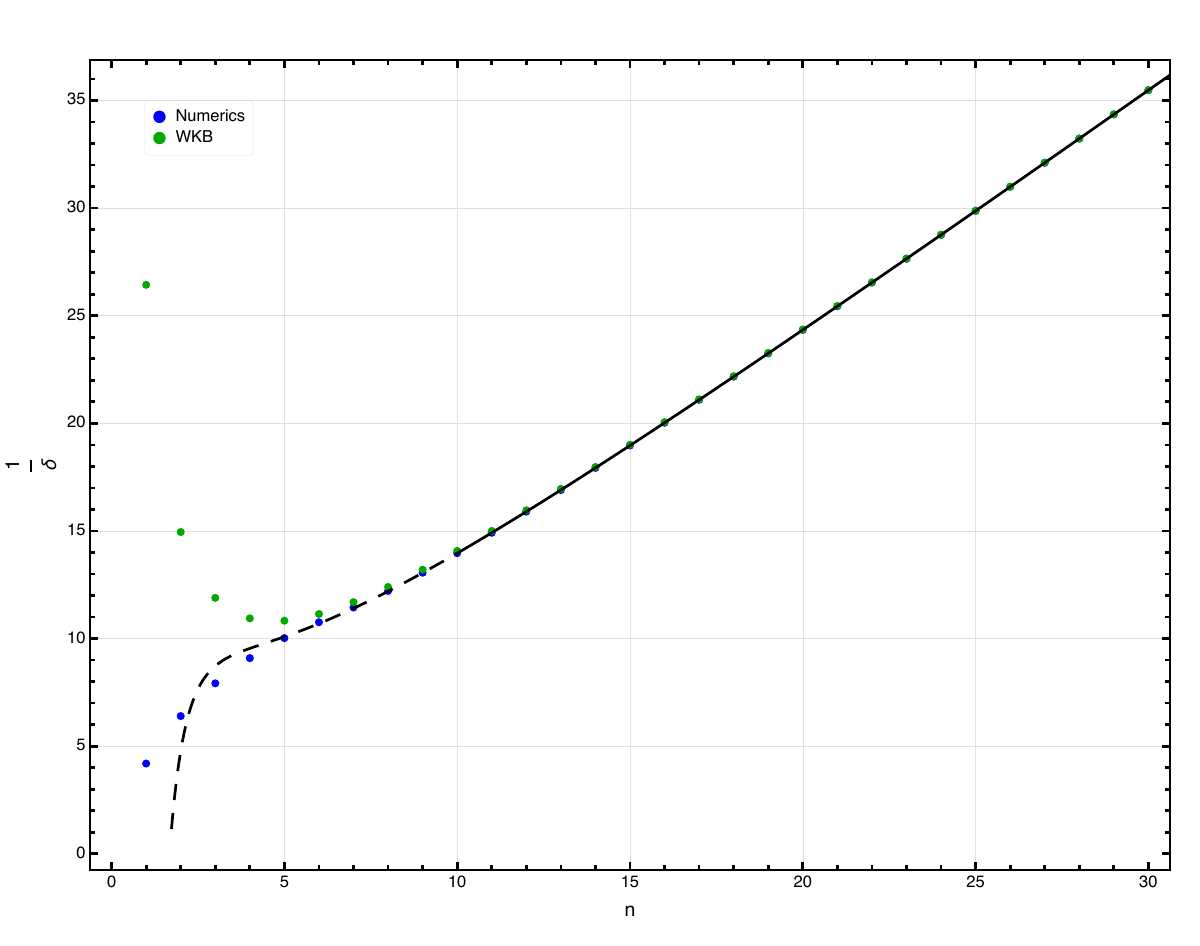}
  \end{minipage}%
  \hfill
  \begin{minipage}[c]{0.30\textwidth}
    \centering
    \begin{tabular}{@{}cccc@{}}
      \toprule
      $n_z$  & $\Delta \alpha$ (\%) \\
      \midrule
  0 & $4.41 \times 10^{-8}$ \\
  1 & $8.81 \times 10^{-7}$ \\
  2 & $1.98 \times 10^{-5}$ \\
  3 & $1.06 \times 10^{-4}$ \\
  4 & $3.41 \times 10^{-4}$ \\
  5 & $8.36 \times 10^{-4}$ \\
      \bottomrule
    \end{tabular}
  \end{minipage}
  \caption{We set $\epsilon=2$ (representative of the experimental aspect ratio). Left: spectrum for $n_z = 3$ with the fitted asymptotic curve (dotted and full black curve) $\frac{1}{\delta}=\frac{q_0}{\omega}=\frac{2}{\sqrt{3}}n+\frac{\alpha}{n}+\frac{\beta}{n^2}+\frac{\gamma}{n^3}+O(\frac{1}{n^4})$. We see that after the first 10 eigenvalues the fit perfectly matches WKB and the numerical eigenvalues.
           Right: difference between the WKB prediction $\alpha_{th}=\frac{-20 + ( \epsilon \pi n_z)^2}{32\sqrt{3}}$  and the numerical prediction. We define $\Delta \alpha= \frac{(\alpha_{num} - \alpha_{th})}{\alpha_{th}}\times 100$. The numerical fit is done between the $100^{th}$ and $200^{th}$ eigenvalues, compared to the theoretical prediction across several
           values of $n_z$. For higher $n_z$ the difference is bigger because the $\frac{1}{n}$ is larger and one needs to add more eigenvalues to obtain a better precision.}
  \label{fig:spectrum-verification}
\end{figure}

\end{itemize}
\subsection{High-energy regime (NLO EFT -- LO WKB)}
In the high-energy regime $\delta \ll \eta \ll 1$, the physical optics approximation 
is valid but NLO corrections to the EFT must be included. The equation of motion is 
now fourth-order in derivatives. As in the spherical case, the angular part contributes 
only at order $\delta^2$ and can be dropped at LO WKB. The contributions to the 
equation of motion from each low energy constant are
\begin{align}
\textbf{$c_0$:}
   & -    \delta^2 \frac{d}{d\rho}\left(-\rho \frac{5}{2}c_0 \mu^{2}(1-\bar V)^{3/2}\frac{d}{d\rho} \pi\right)+\left( \rho \mu^2 \frac{15}{2}c_0 \sqrt{1-\bar V}\right)\pi=0,\\
    \textbf{$c_1$:}&-\delta ^2 \frac{d}{d\rho }\left(\rho  \big(\frac{ \delta^2 q_0^2(\nabla \bar V)^2}{4 (1-\bar V)^{3/2}}+ \frac{2 q_0^2  }{\sqrt{1-\bar V}})\frac{d}{d\rho }\pi \right) + \delta ^2\frac{4   q_0^2}{(1  - \bar V)^{3/4}  }\frac{d}{d\rho  }(\rho  \frac{d}{d\rho }(1 - \bar V)^{1/4}) \pi,  \\
\textbf{$c_2$: }   & \delta^4\frac{d^2}{d \rho^2}\left(-2  q_0^2 c_2 \rho \sqrt{1-\bar{V}} \frac{d^2}{d \rho^2} \pi\right)- \delta^2 \frac{d}{d \rho}\left(\delta^2 q_0^2 c_2\left(\frac{ \bar{V}^{\prime}}{2 \sqrt{1-\bar{V}}}- 2 \frac{\sqrt{1-\bar{V}}}{\rho  }\right)\frac{d}{d \rho} \pi\right)=0.
\end{align}
The equation of motion can be cast into a Sturm--Liouville form,
\begin{equation}
    \delta^4\frac{d^2}{d^2\rho}\left(p(\rho)\frac{d^2}{d^2\rho}\pi \right)-\delta^2\frac{d}{d\rho}\left(s(\rho)\frac{d}{d\rho}\pi \right)=w(\rho)\pi,
\end{equation}
with the following identifications:
\begin{align}
    p(\rho)&=-4 q_0^2 \rho \sqrt{1-\bar{V}}  c_2,\\
    s(\rho)&= \left(-5 \rho  \mu^{2}(1-\bar V)^{3/2}\right)c_0+\left(\rho  \big(\frac{ \delta^2 q_0^2(\nabla \bar V)^2}{2 (1-\bar V)^{3/2}}+ \frac{4 q_0^2  }{\sqrt{1-\bar V}}) \right) c_1+ 2 \delta^2 q_0^2 \left(\frac{ \bar{V}^{\prime}}{2 \sqrt{1-\bar{V}}}- 2 \frac{\sqrt{1-\bar{V}}}{\rho  }\right) c_2,\\
    w(\rho)&=-15 \rho \mu^2 \sqrt{1-\bar V} c_0 -\delta ^2\frac{8   q_0^2}{(1  - \bar V)^{3/4}  }\frac{d}{d\rho}(\rho \frac{d}{d\rho}(1 - \bar V)^{1/4})c_1.
    \end{align}
This is the cylinder analogue of the spherical result, with the geometric 
factor $\rho$ (two-dimensional) replacing $u^2$ (three-dimensional). The eikonal 
and transport equations give
\begin{align}
 &5c_0 \mu^2 (1-\bar V(\rho)) \left(3- (1-\bar V(\rho)) S'_0(\rho)^2\right)+4q_0^2S_0'(\rho)^2(c_1-c_2(1-\bar V(\rho))S_0'(\rho)^2)=0,\\
 &-5c_0\mu^2 (1-\bar V(\rho)) ^2\left(-2(1+2 \rho S_1'(\rho))(1-\bar V(\rho)) +3\rho \bar V'(\rho) -2\rho(1-\bar V(\rho))S_0''(\rho)\right)\nonumber \\
 &\hspace{1cm} 4q_0^2c_1\left(S_0'(\rho)(-2(1+2 \rho S_1'(\rho))(1-\bar V(\rho)) -\rho \bar V''(\rho)-2\rho(1-\bar V(\rho)S_0''(\rho))\right)\nonumber \\
 &\hspace{2cm} 4q_0^2c_2(4(1-\bar V(\rho))S_0'(\rho)^3((1+2 \rho S_1'(\rho))(1-\bar V(\rho)) +\rho \bar V''(\rho) \nonumber\\
 &\hspace{3cm} + 12\rho(1-\bar V(\rho))^2S_0'(\rho)^2S_0''(\rho))=0,
\end{align}
we find the same solutions as in the spherical case,
\begin{align}
  \label{eq:NLO-EFT-LO-WKB}
	S_0 &= \pm \sqrt{3} \int_{\rho_0}^\rho \dd{w} \frac{1}{\sqrt{1-\bar V(w)}}\pqty*{1+\frac{2}{5c_0}\frac{c_1-3c_2}{\pqty*{1-\bar V(w)}^2}\frac{q_0^2}{\mu^2}+\order*{\frac{q_0^4}{\mu^4}} }, \\
	S_1 &= k_1 - \log*(\sqrt{\rho(1-\bar V}) +\frac{(c_1 - 9
    c_2)}{5 c_0(1-\bar V)^2}\frac{q_0^2}{\mu^2}+\order*{\frac{q_0^4}{\mu^4}} .
\end{align}
These are exactly the equations obtained in the high energy regime for a spherically symmetric potential. As expected since we are probing short distances, well below
the scale of the cloud, the global geometry of the trap is irrelevant. The conclusion is the same as Section 4.2 of~\cite{beane2025trappingpotentialdependenceunitaryfermi}. We summarize here the main lesson; adding the NLO effect decreases the size of the droplet. The combination $c_1-9 c_2$ could be positive ( $c_1=-0.0153$, and $c_2=-0.0035$ from mean-field ~~\cite{Ma_es_2009,Kurkjian_2016,Hellerman:2023myh} ) or negative ( $c_1=-0.0209$,
$c_2=\mathcal{O}(\epsilon^2)$ from $\epsilon$- expansion \cite{Rupak_2009}).  The leading order of the spectrum equation receives a correction that scales like $n^{3/2}$ \cite{beane2025trappingpotentialdependenceunitaryfermi}, with a coefficient that is computable in terms of the low-energy constants in the positive case or requires the inclusion of edge terms for the negative case.
\section{Dynamic structure factor}\label{Sec:DSF}

As mentioned in the introduction, what can be accessed in the experiment is the dynamic structure factor from Bragg spectroscopy. The probability per unit time and particle to excite the many-body system from its ground state  by transferring the momentum $q$ and energy $\omega$ is given by Fermi’s golden rule ~\cite{Biss_2022}, $P(q,\omega) = 2\pi \Omega^2_R S(q,\omega)$ with the dynamic structure factor
\begin{equation}
S( q_0,\textbf{q})=\sum_\textbf{n}\left| \langle \textbf{n} | \delta \hat{\rho}^{\dagger}(\textbf{q})| 0  \rangle \right|^ 2 \delta\left(q_0-q_0(\textbf{n})\right), 
\end{equation}
the excited states $|\textbf{n}\rangle$ with energy $q_0(\textbf{n})$, the Fourier transform of the density operator $\delta \hat{\rho}^{\dagger}(\textbf{q})= \sum_{\vec{k}} \hat{a}_{\vec{k}+ \vec{q}}^{\dagger} \hat{a}_{\vec{k}}$ and the Rabi frequency $\Omega_R$ of the atom-light field coupling.
\subsection{ Linear energy regime}
This section is almost identical to the spherical case, we will summarize the procedure.
The Fourier transform of the density, dropping the normalization factor is
\begin{align}
    \braket{0 | \delta \rho(\mathbf{q})| n n_\theta' n_z } &\sim \int \frac{\dd^3{\mathbf{r}}}{(2\pi)^3} e^{-i \textbf{q} \cdot \textbf{r}}e^{i k_z z}e^{i n_\theta' \vartheta} \frac{e^{\frac{i}{\delta} S_0(\rho)   }}{\sqrt{\rho}}\\
    &\sim \int \frac{r d r d\vartheta dz}{(2\pi)^3} e^{-i ( q_z\cdot z+\mathbf{q_\perp \cdot r})}e^{i k_z z}e^{i n_\theta' \vartheta} \frac{e^{\frac{i}{\delta} S_0(\rho)}}{\sqrt{\rho}}
\end{align}
Where we can use the Bessel's first integral to integrate over $\theta$,
\begin{equation}
  J_{n_\theta}(z)=\frac{1}{2 \pi i^{n_\theta}} \int_0^{2 \pi} e^{i z \cos \vartheta} e^{i n_\theta \vartheta} d \vartheta.
\end{equation}
The Fourier transform reduces to the radial integration in the s-wave, 
\begin{align}
 \braket{0 | \delta \rho(\mathbf{q})| n \frac{1}{2} n_z }&\sim \delta(q_z-k_z) \int \frac{r dr }{(2\pi)}J_{1/2}(-q_\perp r)\frac{e^{\frac{i}{\delta} S_0(\rho)}}{\sqrt{\rho}},\\
 &\sim \delta(q_z-k_z)\sqrt{\frac{2}{\pi  q_\perp }} \int \frac{ d \rho  }{(2\pi)}e^{-i R_{cl,\rho }q_\perp \rho }e^{\frac{i}{\delta} S_0(\rho)  },
\end{align}
that can be solved by saddle point. In the longitudinal direction there is translation invariance, the Fourier integral yield delta functions. In the linear regime, we found that $S_0(\rho)$ is given by (\ref{equ:linear-sol}) so the saddle-point equation is
\begin{equation}
    q_{\perp}=\frac{q_0}{\sqrt{2 \mu}} \frac{\sqrt{3}}{\sqrt{1-\bar{V}(\rho)}}=\sqrt{\frac{3}{2(\mu-V(r))}} q_0 .
\end{equation}
This is the curve along which the Fourier transform localizes.
The peaks of the dynamic structure factor appear at the roots of $S_0''(\rho)$. 
\begin{equation}
S_0^{\prime \prime}(\rho)=-\frac{\sqrt{3} \bar{V}^{\prime}(\rho)}{2(1-\bar{V}(\rho))^{3 / 2}}=0 .   
\end{equation}
So we have peaks corresponding to the stationary points of $\bar{V}(\rho)$. For a power-law potential, this means $\mathrm{\rho}=0$. The result is that, at this order, the dynamic structure factor is peaked for

\begin{equation}
   q_{\perp}=\left.\sqrt{\frac{3}{2(\mu-V(\rho))}} q_0\right|_{\rho=0}=\sqrt{\frac{3}{2 \mu}} q_0, 
\end{equation}

which is independent of the potential (that vanishes at the origin). As expected, the leading behavior of the spectrum is linear, since the fluctuations around the ground state are described by a type-I Goldstone.
\subsection{ Low energy regime}
We will focus on the dynamic structure factor in the low energy regime, as the discussion in the high energy regime is identical to the spherical case for $q=q_\perp$ since the longitudinal modes are fixed by $q_z=\frac{\pi n_z}{L}$.
Dropping overall constant factors, we reduce the Fourier transform of the density 
fluctuation to a single $\rho$-integral that we can solve by saddle point
\begin{align}
    \braket{0 | \delta \rho(\mathbf{q})| n n_\theta' n_z } &\sim \int \frac{\dd^3{\mathbf{r}}}{(2\pi)^3} e^{-i \textbf{q} \cdot \textbf{r}}e^{i k_z z}e^{i n_\vartheta' \vartheta} \frac{e^{\frac{i}{\delta} S_0(\rho)  + i \delta S_2(\rho) }}{\sqrt{\rho}},\\
     &\sim \delta(q_z-k_z)\sqrt{\frac{2}{\pi  q_\perp }} \int \frac{ d \rho  }{(2\pi)}e^{-i R_{cl,\rho }q_\perp \rho }e^{\frac{i}{\delta} S_0(\rho)  + i \delta S_2(\rho) }.
\end{align}
The saddle condition on the resulting integral is
\begin{equation}
    \frac{d}{d\rho}\left( R_{cl,\rho} q_\perp \rho \delta -S_0(\rho)   - \delta^2 S_2(\rho)\right)=0,
\end{equation}
which, upon restoring units, gives the curve where the expectation value of the density fluctuations localizes 
\begin{equation}
    q_\perp=\sqrt{\frac{3}{2(\mu-V(r))}}q_0\left[1+\frac{1}{6q_0^2}\left(V''(r)+\frac{3V'(r)}{2r}-\frac{2(\pi n_z)^2}{L^2}(\mu-V(r))\right)\right].
\end{equation}
The peaks of the dynamic structure factor appear at the roots of the equation
\begin{equation}
\odv{}{r} \bqty*{ \sqrt{\frac{3}{2(\mu-V(r))}}q_0\left[1+\frac{1}{6q_0^2}\left(V''(r)+\frac{3V'(r)}{2r}-\frac{2(\pi n_z)^2}{L^2}(\mu-V(r))\right)\right] } = 0.
\end{equation}
In the special case of the harmonic potential $\varpi=\omega$, this equation admits the solutions $r=0$. The dynamic structure factor peaks along
\begin{align} \label{equ:dispersion-relation-LO}
q_0(\mathbf{q})=\sqrt{\frac{2\mu}{3}} \mathrm{q_\perp} \left(1+\left(\frac{ n_z^2 \pi^2}{2\mathrm{~L}^2 }-\frac{5 \omega^2}{8  \mu}\right)\frac{1}{q_\perp^2}\right),\\
q_0(\mathbf{q})=\sqrt{\frac{2\mu}{3}} \mathrm{q_\perp} \left(1+\left(\frac{n_z^2 \pi^2}{2\mathrm{~L}^2 }-\frac{5}{4 R^2_{cl,\rho}}\right)\frac{1}{q_\perp^2}\right).
\end{align}
The longitudinal confinement always contributes a convex correction, while the 
transverse trap gives a negative (concave) correction. The curvature depends on the size of the cylinder at a given longitudinal mode. For $n_z\geq1$, given the experimental values $R=60 \mu m,L=50 \mu m$~\cite{Patel_2020}, $R=50\mu m, L=43\mu m$ ~\cite{Biss_2022},
the combination $\frac{ n_z^2 \pi^2}{2 L^2}-\frac{5}{4 R^2}>0$ is always positive.
For more general transverse potentials, the dynamic structure factor peaks remain at $r=0$, the center of the trap, for flatter transverse potentials 
the negative correction coming from the potential is further suppressed, following the pattern observed in the 
spherical case.\\

\subsection{Comparison with experiment}
As mentioned in Section \ref{Sec:Experiments}, the experimental analysis assumes complete isotropy. However, even in the presence of a very steep transverse confinement, this potential explicitly breaks isotropy.
In our framework, the dynamic structure factor depends on the full wavevector $\mathbf{q}$ rather than only its magnitude $q$, (\ref{equ:dispersion-relation-LO}) can be written explicitly, using $q_z=\frac{n_z \pi}{L}$, as,
\begin{equation}
    q_0(\mathbf{q})=\sqrt{\frac{2\mu}{3}} \mathrm{q_\perp} \left(1+\left(\frac{q_z^2}{2 }-\frac{5 \omega^2}{8  \mu}\right)\frac{1}{q_\perp^2}\right).
\end{equation} 
In \cite{Biss_2022}, the perturbation induced by the Bragg laser is purely radial, our analysis is therefore strictly applicable in the case $n_z = 0$. In our setup, this configuration corresponds to the closest realization of an effectively isotropic system.
Under these conditions, (\ref{equ:dispersion-relation-LO}) reduces, in the case of a transverse harmonic potential, to
\begin{align} 
q_0(q)=\sqrt{\frac{2\mu}{3}} \mathrm{q} \left(1-\frac{5 \omega^2}{8  \mu}\frac{1}{q^2}\right).
\end{align}
In the case of a flatter potential, $V(\rho)=\rho^{16}$,

\begin{align} \label{equ:disperion_flatter_potential}
q_0(q)=\sqrt{\frac{2\mu}{3}} \mathrm{q} \left(1-O\left(\frac{\varpi^{16}}{q^{16} \mu^8}\right)\right).
\end{align}
Comparing our results with experiment, we expect a correction to the linear behavior, always convex in the high-energy regime. In the low-energy regime, the correction is convex except in the case $n_z=0$, where only radial modes are excited, the correction is concave (\ref{equ:disperion_flatter_potential}).
\section{Conclusion}

We have extended the EFT treatment of~\cite{beane2025trappingpotentialdependenceunitaryfermi} to a cylindrical geometry, 
which more closely reflects current experimental realizations of the unitary Fermi gas~\cite{Patel_2020}. 
The cylinder introduces a new geometric scale absent in the spherically symmetric case: the 
longitudinal half-length $R_{\mathrm{cl},z} = L/2$, which enters the dynamic structure factor
alongside the transverse cloud radius $R_{\mathrm{cl},\rho}$. The ratio $\epsilon = 
R_{\mathrm{cl},\rho}/R_{\mathrm{cl},z}$ characterizes the aspect ratio of the trap, we made no assumption on the magnitude, keeping the analysis valid for the nearly cubic 
geometries used in recent experiments~\cite{Biss_2022, Patel_2020}.

In the high-energy regime ($\delta \ll \eta \ll 1$), where the physical optics approximation 
is valid and NLO EFT corrections are required, we find that the dynamic structure factor is 
insensitive to the trapping geometry: the corrections depend only on the low energy constant
$c_1$ and $c_2$, and the result is structurally identical 
to the spherical case, predicting a convex correction. This is physically transparent --- the high-energy regime probes 
short distances, well below the scale of the cloud, so the global geometry of the trap is 
irrelevant and only the local, microscopic physics encoded in the low-energy coefficients matter.

In the low-energy regime ($\eta \ll \delta \ll 1$), by contrast, the geometry plays a central role. 
The most important effect appears for the harmonic potential, where the NLO WKB correction to the dynamic structure factor, depends on the combination,
\begin{equation}
    \frac{n_z^2\pi^2}{2 L^2} - \frac{5}{4R_{\mathrm{cl},\rho}^2},
\end{equation}
where the longitudinal confinement always contributes a convex correction and the 
transverse trap a concave one. Evaluating this combination for the geometries used in  current experiments we find that it is positive in all cases, where $n_z\ne0$, so that the net correction is convex. 

This provides a concrete, geometry-dependent prediction that can be tested against Bragg spectroscopy  measurements of the dynamic structure factor.
These techniques are thus applicable to different geometries. Another interesting case to study would be the ellipsoid, for which the spectrum has already been computed in \cite{Csord_s_2000}.

\section*{Acknowledgments}

The author would like to thank Yvan Castin for valuable comments on the draft as well as Silas Beane, Domenico Orlando, and Susanne Reffert for  enlightening discussions and their comments on the draft. This work is supported by the Swiss National Science Foundation under grant No. 200021\_219267.

\setstretch{1}
\small\sffamily

\bibliography{bibi}{}

\bibliographystyle{JHEP}
\end{document}